\documentclass{jfm}
\usepackage{amssymb,latexsym,bm,amsmath,epsfig}

\usepackage[OT1]{fontenc}
\usepackage[applemac]{inputenc}
\catcode`\ú=\active \gdefú{\setbox0=\hbox{0}\hbox to\wd0{}}

\ifx\pdfoutput\defined
\usepackage{graphicx}
\graphicspath{{figuresEPS/}}
\else
\usepackage[pdftex]{graphicx}
\usepackage{epstopdf}
\graphicspath{{figuresEPS/}}
\fi
\title[Acceleration statistics of heavy particles in
turbulence]{Acceleration statistics of heavy particles in turbulence}

\author[J.~Bec {\it et al.}]%
{J.\ns B\ls E\ls C$^{\,1}$,\ns
L.\ns B\ls I\ls F\ls E\ls R\ls A\ls L\ls E$^{\,2}$,\ns
G.\ns B\ls O\ls F\ls F\ls E\ls T\ls T\ls A$^{\,3}$,
\ns
A.\ns C\ls E\ls L\ls A\ls N\ls I$^{\,4}$,\ns
M.\ns C\ls E\ls N\ls C\ls I\ls N\ls I$^{\, 5}$,\ns
A.\ns L\ls A\ls N\ls O\ls T\ls T\ls E$^{\, 6}$,
\ns
S.\ns M\ls U\ls S\ls A\ls C\ls C\ls H\ls I\ls O$^7$,\ns 
\and \ls F.\ns T\ls O\ls S\ls C\ls H\ls I$^{8}$ }

\affiliation{$^1$ CNRS Observatoire de la C\^ote d'Azur, B.P. 4229,
06304 Nice Cedex 4, France\\ [\affilskip] $^2$ Dept. of Physics and
INFN, University of Rome ``Tor Vergata'', \\Via della Ricerca
Scientifica 1, 00133 Roma, Italy\\ [\affilskip] $^3$ Dept. of Physics
and INFN, University of Torino, Via Pietro Giuria 1, 10125, Torino,
Italy\\ [\affilskip] $^4$ CNRS, INLN, 1361 Route des Lucioles,
F-06560 Valbonne, France\\ [\affilskip] $^5$ SMC-INFM c/o Dept. of
Physics University of Rome ``La Sapienza", Piazz.le A.~Moro, 2,
I-00185 Roma, Italy, and CNR-ISC via dei Taurini, 19 I-00185 Roma,
Italy \\ [\affilskip] $^6$ CNR-ISAC, Sezione di Lecce,
Str. Prov. Lecce-Monteroni km 1,200, I-73100 Lecce, Italy\\ [\affilskip]
$^7$ Dept. of Physics, University of Rome ``La Sapienza", Piazz.le
A.~Moro, 2, I-00185 Roma, Italy\\ [\affilskip] $^8$ CNR-IAC, Viale del
Policlinico 137, I-00161 Roma, Italy and\\ INFN, Sezione di Ferrara, via
G. Saragat 1, I-44100, Ferrara, Italy}

\pubyear{????}
\volume{???}
\pagerange{??--???}
\begin{document}

\maketitle

\begin{abstract}
We present the results of direct numerical simulations of heavy
particle transport in homogeneous, isotropic, fully developed
turbulence, up to resolution $512^3$ ($R_\lambda\approx 185$).
Following the trajectories of up to 120 million particles with Stokes
numbers, $St$, in the range from $0.16$ to $3.5$ we are able to
characterize in full detail the statistics of particle
acceleration. We show that: ({\it i}\/) The root-mean-squared
acceleration $a_{\rm rms}$ sharply falls off from the fluid tracer
value already at quite small Stokes numbers; ({\it ii}\/) At a given
$St$ the normalised acceleration $a_{\rm rms}/(\epsilon^3/\nu)^{1/4}$
increases with $R_\lambda$ consistently with the trend observed for
fluid tracers; ({\it iii}\/) The tails of the probability density
function of the normalised acceleration $a/a_{\rm rms}$ decrease with
$St$.  Two concurrent mechanisms lead to the above results: 
preferential concentration of particles, very effective at small $St$,
and filtering induced by the particle response time, that takes over
at larger $St$.
\end{abstract}
\section{Introduction}
Small impurities like dust, droplets or bubbles suspended in an
incompressible flow are finite-size particles whose density may differ
from that of the underlying fluid, and cannot thus be modelled as
point-like tracers. The description of their motion must account for
inertia whence the name {\em inertial particles}.  At long times
particles concentrate on singular sets evolving with the fluid motion,
leading to the apparition of a strong spatial inhomogeneity dubbed
{\em preferential concentration}.  At the
experimental level such inhomogeneities have been long known (see
\cite{3} for a review) and utilised for flow visualisation
(e.g. exploiting bubble clustering inside vortex filaments). The
statistical description of particle concentration is at present a
largely open question with many industrial and environmental
applications. We mention spray combustion in Diesel engines (\cite{4})
or some rocket propellers (\cite{prop}), the formation of rain droplets
in warm clouds (\cite{5,7,8}) or the coexistence of plankton species
(\cite{9,10}).  Inertial particles are also relevant to spore, pollen,
dust or chemicals dispersion in the atmosphere where the diffusion by
air turbulence may be even overcome by preferential clustering
(\cite{11,12}).

From the experimental side, the study of particle motion in turbulence
has recently undergone rapid progress thanks to the development of
effective optical and acoustical tracking techniques
(\cite{boden,boden1,Pinton,wara}). In parallel with experimental effort,
theoretical analysis
(\cite{Falkovich-clustering,Falkovich-Pumir,Bec-Gaw-Horvai,simo}) and
numerical simulations (\cite{Squires,Collins,Wang,Collins2}) are paving the way
to a thorough understanding of inertial particle dynamics in turbulent
flows.  Recently, the presence of strong inhomogeneities characterised
by fractal and multifractal properties have been predicted, and found
in theoretical and numerical studies of stochastic laminar flows
(\cite{Falkovich-clustering,Bec-Gaw-Horvai,Bec}), in two dimensional
turbulent flows (\cite{Boffetta}) and in three dimensional turbulent
flows at moderate Reynolds numbers in the limit of vanishing inertia
(\cite{Falkovich-Pumir}).

Here we present a Direct Numerical Simulations (DNS) study of
particles much heavier than the carrier fluid in high-resolution
turbulent flows.  In particular, we shall focus on the behaviour of
particle acceleration at varying both Stokes and Reynolds numbers.
For fluid tracers, it is known that trapping into vortex filaments
(\cite{boden,bife}) is the main source of strong acceleration
events. On the other hand, little is known about the acceleration
statistics of heavy particles in turbulent flows, where preferential
concentration may play a crucial role.  Moreover, since in most
applied cases it is almost impossible to perform DNS of particle
transport in realistic settings, it is very important to understand
acceleration statistics for building stochastic models of particle
motion with and without inertia (\cite{pdfmod}).
 
The material is organised as follows.  In~\S\,2, we briefly recall the
equations of motion of the inertial particles and summarise the DNS
set up.  In~\S\,3, we present and discuss the main results concerning
the acceleration statistics.  In~\S\,4 we discuss the acceleration
statistics conditioned on the local structure of the flow. The final
Section is devoted to conclusions and perspectives.

\section{Heavy particle dynamics and numerical simulations}
The equations of motion of a small, rigid, spherical particle immersed
  in an incompressible flow have been consistently derived from first
  principles by \cite{maxey}. In the limiting case of particles much
  heavier than the surrounding fluid, these equations take the
  particularly simple form
\begin{equation}
{d {\bm X} \over dt} = {\bm V}(t) \; , \qquad
{d {\bm V} \over dt} = -\frac{{\bm V}(t)-{\bm u}({\bm X}(t),t)}{\tau_s} \;. 
\label{eq:1}
\end{equation}
Here, ${\bm X}(t)$ denotes the particle trajectory, ${\bm V}(t)$ its
velocity, ${\bm u}({\bm x},t)$ is the fluid velocity. The Stokes
response time is $\tau_s= 2 \rho_p a^2 /(9 \rho_f \nu)$ where $a$ is
the particle radius $\rho_p$ and $\rho_f$ are the particle and fluid
density, respectively, and $\nu$ is the fluid kinematical viscosity.
The Stokes number is defined as $St=\tau_s/\tau_\eta$ where
$\tau_\eta=(\nu/\epsilon)^{1/2}$ is the Kolmogorov timescale and
$\epsilon$ the average rate of energy injection.  Eq.~(\ref{eq:1}) is
valid for very dilute suspensions, where particle-particle
interactions (collisions) and hydrodynamic coupling are not taken into
account.

The fluid evolves according to the incompressible Navier-Stokes
equations
\begin{equation}
  \frac{\partial{\bm u}}{\partial t} + {\bm u} \cdot {\bm \nabla} {\bm u} =
  -\frac{{\bm \nabla} p}{\rho_f} + \nu \Delta {\bm u} + {\bm f}\,,
\label{eq:2}
\end{equation}
where $p$ is the pressure field and ${\bm f}$ is the external energy
source, $\langle {\bm f}\cdot{\bm u} \rangle = \epsilon$.

The Navier-Stokes equations are solved on a cubic grid of size $N^3$
for $N=128,\,256,\,512$ with periodic boundary conditions.  Energy is
injected by keeping constant the spectral content of the two smallest
wavenumber shells (\cite{She}).  The viscosity is chosen so to have a
Kolmogorov lengthscale $\eta\approx \Delta x$ where $\Delta x$ is the
grid spacing: this choice ensures a good resolution of the small-scale
velocity dynamics. We use a fully dealiased pseudospectral algorithm
with 2$^{nd}$ order Adam-Bashforth time-stepping. The Reynolds numbers
achieved are in the range $R_{\lambda} \in [65: 185]$.  

The equations of fluid motion are integrated until the system reaches
a statistically steady state.  Then, particles are seeded with
homogeneously distributed initial positions and velocities equal to
the local fluid velocity.  Equations (\ref{eq:1}) and (\ref{eq:2}) are
then advanced in parallel.  A transient in particle dynamics follows,
about $2-3$ large scale eddy turn over time, before reaching a
Lagrangian stationary statistics.  It is only after this relaxation
stage has completely elapsed that the real measurement starts. We
followed $15$ sets of inertial particles with Stokes numbers from
$0.16$ to $3.5$. For each set, we saved the position and the
velocity of $N_t$ particles every $dt = 1/10 \tau_{\eta}$ with a
maximum number of recorded trajectories of $N_t = 5 \cdot 10^{5}$ for
the highest resolution. Along these trajectories we also stored the
velocity of the carrier fluid.  At a lower frequency $\sim 10
\tau_\eta$, we saved the positions and velocities of a larger number
$N_p$ of particles (up to $7.5\cdot 10^6$ per $St$ at the highest
resolution) together with the Eulerian velocity field.  We have also
followed fluid tracers ($St=0$), that evolve according to the dynamics
\begin{equation}
\frac{d\bm x(t)}{dt} = \bm u (\bm x(t),t)\,,
\label{eq:tracer}
\end{equation}
 in order to
systematically assess the importance of the phenomenon of preferential
concentration at varying both $St$ and $R_\lambda$.

A summary of the various physical parameters is given in
table~\ref{table:1}.
\begin{table*}
\begin{tabular}{ccccccccccccccc}
  \hline 
  $R_\lambda$ & $u_{\rm rms}$ & $\varepsilon$ & $\nu$ & $\eta$ & $L$ & $T_E$
  & $\tau_\eta$ & $T_{tot}$ & $T_{tr}$ & $\Delta x$ & $N^3$ & $N_t$ & $N_p$ & $N_{tot}$\\ 
  \hline 
  185 & 1.4 & 0.94 & 0.00205 &0.010 & $\pi$ & 2.2 & 0.047 &
  14 & 4 & 0.012 & $512^3$ & $5\cdot 10^{5}$ & $7.5\cdot 10^6$ &$12\cdot 10^7$\\
  105 & 1.4 & 0.93 & 0.00520 &0.020 & $\pi$ & 2.2 & 0.073 & 20 & 4 & 0.024 & $256^3$ & 2.5$\cdot 10^{5}$ & $2\cdot 10^6$ &$32\cdot 10^6$\\
   65 & 1.4 & 0.85 & 0.01  &0.034 & $\pi$ & 2.2 & 0.110 & 29 & 6 & 0.048 & $128^3$ & 3.1$\cdot 10^{4}$ & $2.5\cdot 10^5$ & $4\cdot 10^6$\\
  \hline 
\end{tabular}
\caption{Parameters of DNS. Microscale Reynolds number $R_\lambda$,
root-mean-square velocity $u_{\rm rms}$, energy dissipation
$\varepsilon$, viscosity $\nu$, Kolmogorov lengthscale
$\eta=(\nu^3/\varepsilon)^{1/4}$, integral scale $L$, large-eddy
Eulerian turnover time $T_E = L/u_{\rm rms}$, Kolmogorov timescale
$\tau_\eta$, total integration time $T_{tot}$, duration of the
transient regime $T_{tr}$, grid spacing $\Delta x$, resolution $N^3$,
number of trajectories of inertial particles for each Stokes $N_t$
saved at frequency $\tau_{\eta}/10$, number of particles $N_p$ per
Stokes stored at frequency $10 \tau_\eta$, total number of advected
particles $N_{tot}$. Errors on all statistically fluctuating
quantities are of the order of $10 \%$.}
\label{table:1}
\end{table*}
\section{Results and discussion}
 \begin{figure}
\centering
\includegraphics[draft=false,scale=0.51]{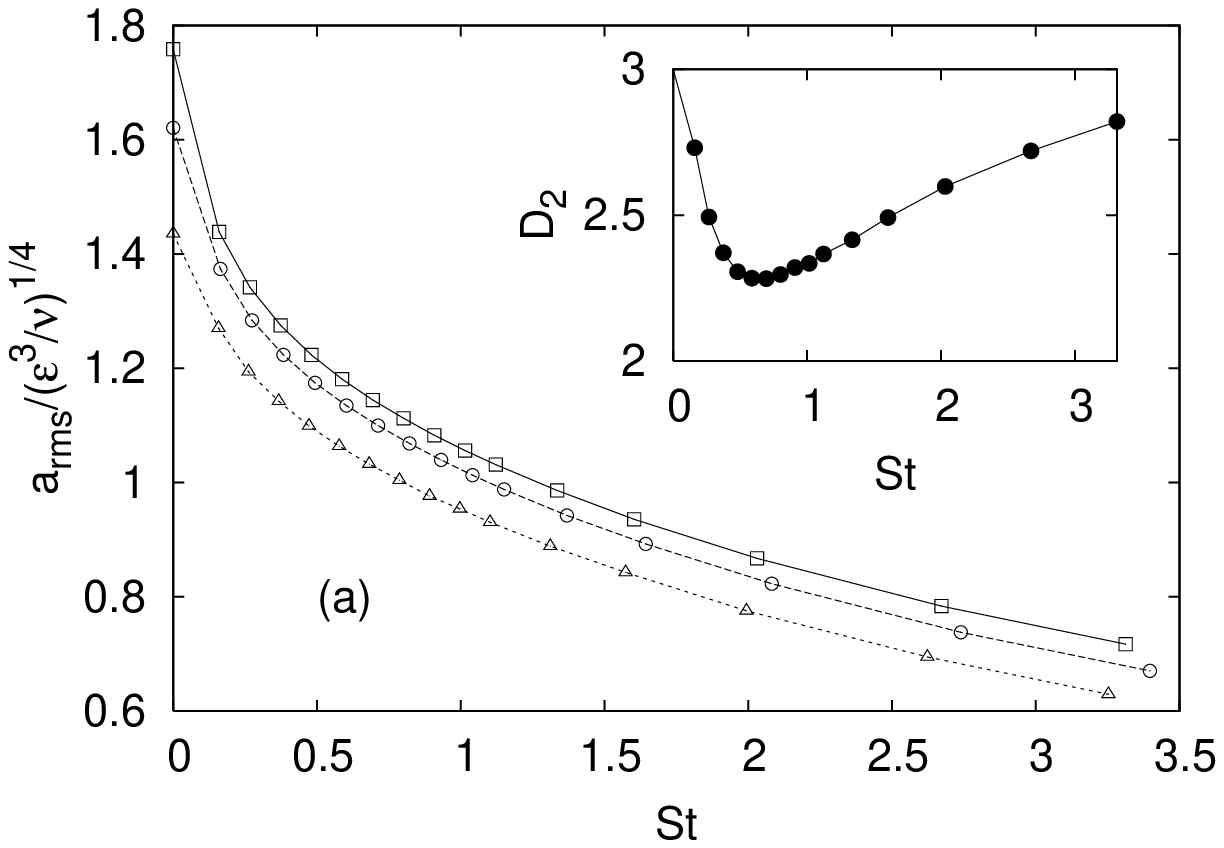} \hfill
\includegraphics[draft=false,scale=0.51]{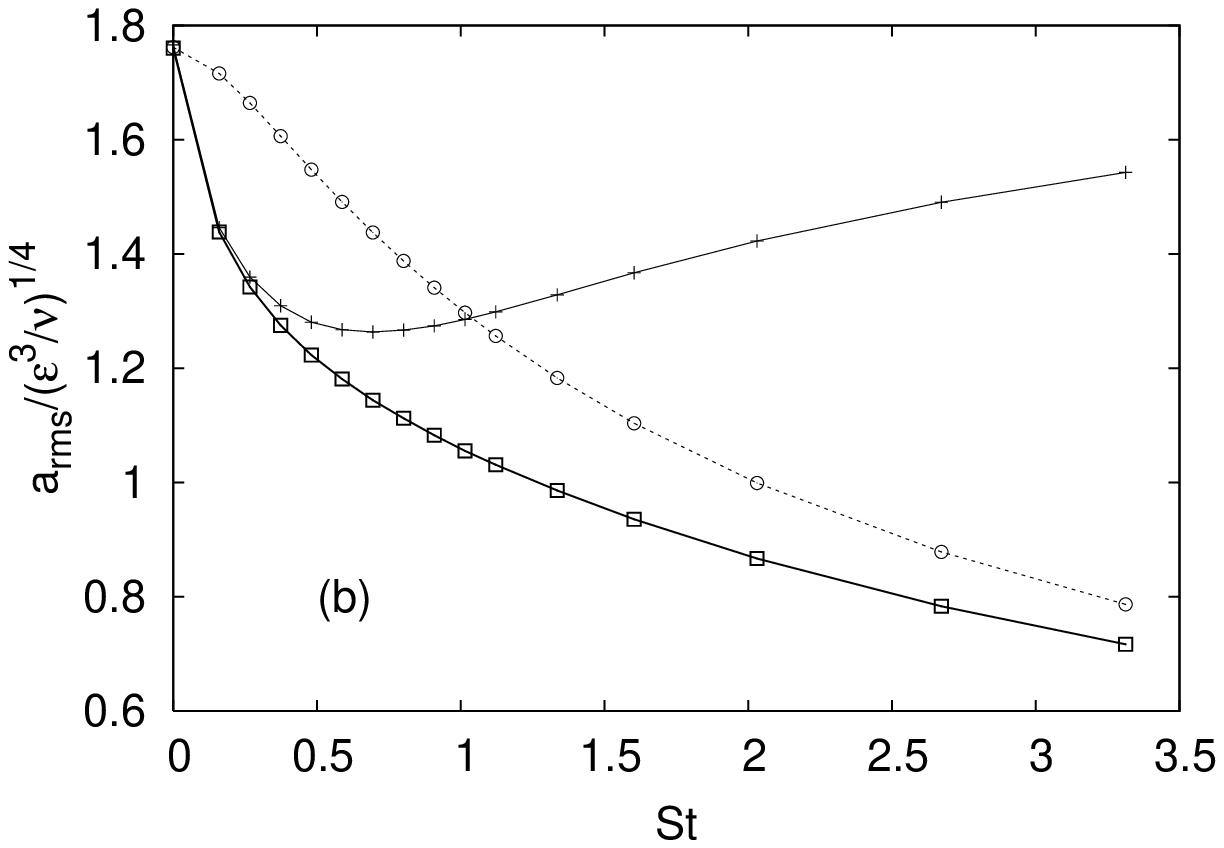}
\caption{ (a) The normalised acceleration variance $a_{\rm
rms}/(\epsilon^3/\nu)^{1/4}$ as a function of the Stokes number for
$R_{\lambda}=185$ ($\Box$); $R_{\lambda}=105$ ($\circ$);
$R_{\lambda}=65$ ($\bigtriangleup$).  The inhomogeneous distribution
of particle is quantified for the highest Reynolds in the inset, where
we plot the correlation dimension, $D_2$, as a function of $St$. The
correlation dimension is defined as $p(r)\sim r^{D_2}$ (for $r \ll
\eta$) where $p(r)$ is the probability to find two particles at
distance smaller than $r$ (\cite{bccm05}).  (b) Comparison between the
acceleration variance, $a_{\rm rms}$ ($\Box$), as a function of Stokes, with
the acceleration of the fluid tracer measured on the particle
position, $\langle ({ D{\bm u}\over Dt})^2 \rangle^{1/2}$ ($+$).
The last curve ($\circ$), approaching the $a_{\rm rms}$ for large $St$, is
the one obtained from the filtered tracer trajectories,
$a^F_{\rm rms}$. All data refer to $Re_{\lambda}=185$.}
\label{fig:1}
\end{figure}

In this paper we focus on the statistics of particle acceleration
${\bm a}(t)={d{\bm V} \over dt}$. From previous studies on fluid tracers
we know that acceleration statistics is very intermittent and strong 
fluctuations are associated to trapping events within vortex filaments
(\cite{boden,boden1,Pinton,bife}). How does inertia
impacts acceleration statistics? A good starting point to gain insight
on the effect of inertia is given by the formal solution of Eqs.~(\ref{eq:1})
in the statistically stationary state, relating the instantaneous 
particle velocity to the previous history of fluid velocity along the
particle trajectory. The expression is 
\begin{equation}
{\bm V}(t)=\frac{1}{\tau_s}\int_{-\infty}^t e^{-(t-s)/\tau_s} {\bm u}({\bm X}(s),s) \, ds
\label{eq:3}
\end{equation}
yielding for the acceleration
 \begin{equation}
{\bm a}(t)=\frac{1}{\tau_s^2}\int_{-\infty}^t e^{-(t-s)/\tau_s} 
[{\bm u}({\bm X}(t),t)-{\bm u}({\bm X}(s),s)]\, ds \;.
\label{eq:4}
\end{equation}
It is instructive to analyse separately the two limiting cases of small
and large Stokes numbers.

\noindent {\em At small $St$}, i.e. $\tau_s \ll \tau_\eta$, the fluid
velocity along the trajectory evolves smoothly in time and the above
expression for the acceleration reduces to ${\bm a(t)} \simeq {d \over
dt} {\bm u}({\bm X}(t),t)$, i.e. to the derivative of fluid velocity
along the inertial particle trajectory. At sufficiently small $St$
this is indistinguishable from the fluid acceleration ${D {\bm u}\over
Dt}({\bm X}(t),t)$ evaluated at particle positions. The latter, in
turn, is essentially dominated by the $-{\bm \nabla}p$
contribution. Therefore we are led to draw the following picture for
the small $St$ case: the heavy particle acceleration essentially
coincides with the fluid acceleration; however, inertial particles are
not homogeneously distributed in the flow and concentrate
preferentially inside regions with relatively small pressure gradient
(low vorticity regions).  As a result, the net
effect of inertia is a drastic reduction of the root-mean-squared
acceleration $a_{\rm rms}= \langle {\bm a}^2 \rangle^{1/2}$, due
essentially to preferential concentration.  Indeed, as shown in
Fig.~\ref{fig:1}a the acceleration variance drops off very fast
already at quite small $St$ values. In Fig.~\ref{fig:1}b we give
evidence that the value of $a_{\rm rms}$ is very close for $St<0.4$ to
$\langle ({ D{\bm u}\over Dt})^2 \rangle^{1/2}$ when the average is
not taken homogeneously in space but conditioned to be on the same
spatial positions of the inertial particles.  The agreement of the two
curves supports the arguments above.  Notice that at increasing Stokes
the two curves start to deviate from each other, the tracer
acceleration conditioned on the particle positions has a minimum for
$St \approx 0.5$ close to the maximum of clustering (see inset of
Fig.~\ref{fig:1}a), eventually recovering the value of $a_{\rm rms}$ of the
unconditioned tracers for larger $St$. The latter effect is a clear
indication that inertial particles explore the small scale structures
of the flow more and more homogeneously by increasing $St$. In this
limit a different mechanism is responsible for the reduction of the
$a_{\rm rms}$.\\
\noindent 
{\em At large $St$}, i.e. $\tau_s \gg \tau_\eta$, the inspection of
Eq.~(\ref{eq:4}) shows that the main effect of inertia on particle
acceleration is a low-pass filtering of fluid velocity differences,
with a suppression of fast frequencies above $\tau_s^{-1}$. In
figure~\ref{fig:1}b we also compare the acceleration variance with the
one obtained by an artificial low-pass filtering based only on the
fluid tracers trajectories.  For each tracer
trajectory, $\bm x(t)$, we define  a new velocity, $\bm u^F$, 
filtered on a window-size of
the same order of the Stokes time:
\begin{equation}
\label{filt}
{\bm u^F}(t)=\frac{1}{\tau_s}\int_{-\infty}^t e^{-(t-s)/\tau_s} {\bm
u}({\bm x}(s),s) \, ds
\end{equation}
The filtered acceleration is thus given by $ {\bm a}^F = \frac{d}{dt}
{\bm u^F}$. Of course, in order to extract the effect due to filtering
only we are compelled to employ fluid trajectories: (\ref{filt})
applied along particle trajectories is nothing but Eq.~(\ref{eq:3}),
so that the acceleration would coincide with the particle acceleration
by definition. The root mean square fluctuation, $a^F_{\rm rms} =
\langle (\frac{d}{dt} {\bm u^F})^2 \rangle^{1/2}$, is thus computed by
averaging along the tracer trajectories without any condition on their
spatial positions, i.e. homogeneously distributed in the whole $3d$
domain.  The curves corresponding to $a_{\rm rms}$ and to $ a^F_{\rm
rms}$ become closer and closer as $St$ grows larger, supporting the conjecture
that preferential concentration for $St>1 $ becomes less
important. For intermediate $St$ we expect a non trivial interplay
between the two above mechanisms that makes very difficult to build up
a model able to reproduce even the qualitative behaviour.
\begin{table*}
\centering
\begin{tabular}{c|ccccccccccccccccc}
\hline
St$^{(a)}$  & 0 & 0.16 & 0.27 &0.37 &0.48 &0.59 &0.69 &0.80 &0.91 &1.01 &1.12 &1.34 &1.60 &2.03 &2.67 &3.31 \\ 
$\langle \tilde{a}^2 \rangle $ & 3.09 & 2.07 & 1.80 & 1.63 & 1.50 &1.39 &1.31 & 1.24 & 1.17 & 1.12 & 1.06 &  0.97 &  0.88 & 0.75 & 0.61 &0.51 \\
$\langle \tilde{a}^4 \rangle $ & 288 & 48.1 & 30.5 & 22.4 & 17.7 & 14.5 & 12.3 & 10.6 & 9.20 & 8.11 &
7.21& 5.77& 4.47	&3.11 &1.94 &1.29\\\hline
St$^{(b)}$  &0 & 0.16 & 0.27 & 0.38 & 0.49 & 0.60 & 0.71 & 0.82 & 0.93 & 1.04 & 1.15 & 1.37 & 1.64 & 2.08 & 2.74 &3.40 \\
$\langle \tilde{a}^2 \rangle $ & 2.63 & 1.89 & 1.65 & 1.45 & 1.38 & 1.29 & 1.21 & 1.14 & 1.08 & 1.03 & 0.98 & 0.89 & 0.80 & 0.68 & 0.54 & 0.45\\
$\langle \tilde{a}^4 \rangle $ &133 & 32.9 & 21.6& 16.3 & 13.1& 10.9& 9.29& 8.03& 7.01& 6.18&5.48&
4.37 & 3.36 &2.23&  1.39 & 0.90\\\hline
St$^{(c)}$ & 0  &  0.16 & 0.26 & 0.37 & 0.47 & 0.58 & 0.68 & 0.79 & 0.89 & 1.00 & 1.10 &  1.31&   1.57 &  1.99 &  2.62 & 3.25 \\
$\langle \tilde{a}^2 \rangle$ & 2.02 &  1.59 &  1.40 &  1.28 &  1.19 &  1.11 &  1.05 &  0.99 &  0.94 &  0.89 & 0.85 &  0.77 &  0.70 &  0.59 &  0.47 &  0.39 \\
$\langle \tilde{a}^4 \rangle$ & 52.8 & 19.1 & 13.1 & 10.1 & 8.24 & 6.95 & 6.01 & 5.24 & 4.61 & 4.11
& 3.67 & 2.95 & 2.32 & 1.59&    0.97 & 0.63\\\hline
\end{tabular}
\caption{Normalised values of the second and fourth moments of the
  acceleration $\langle \tilde{a}^2 \rangle = \langle \bm a^2 \rangle
  / [3 (\epsilon^3/\nu)^{1/2}]$ , $\langle \tilde{a}^4 \rangle =
  \langle \bm a^4 \rangle / [3 (\epsilon^3/\nu)]$ for $^{(a)}$
  $R_\lambda=185$, $^{(b)}$ $R_\lambda=105$ and $^{(c)}$
  $R_\lambda=65$. The statistical error on all entries are of the
  order of $5\%$. }
\label{table:2}
\end{table*}
\begin{figure}
\centering
\includegraphics[draft=false,scale=0.75]{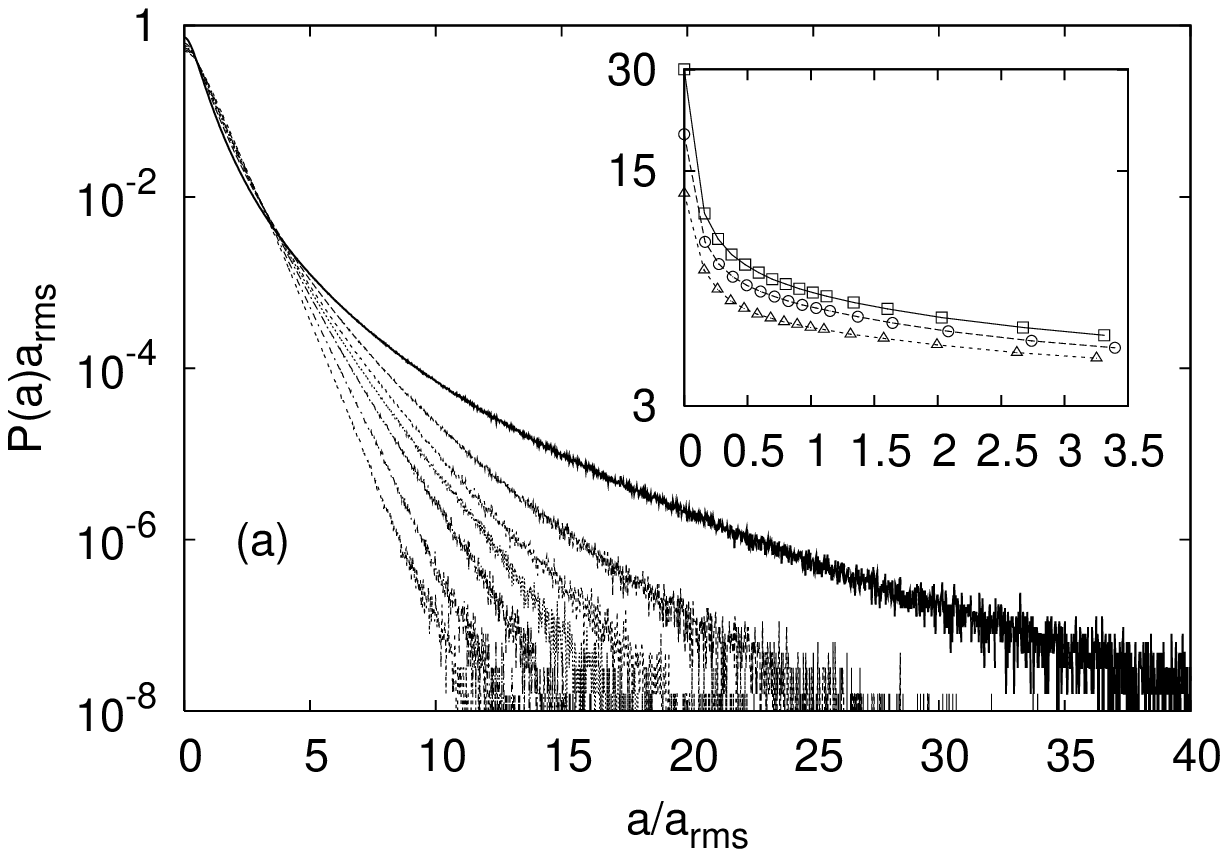}
\centering
\includegraphics[draft=false,scale=0.75]{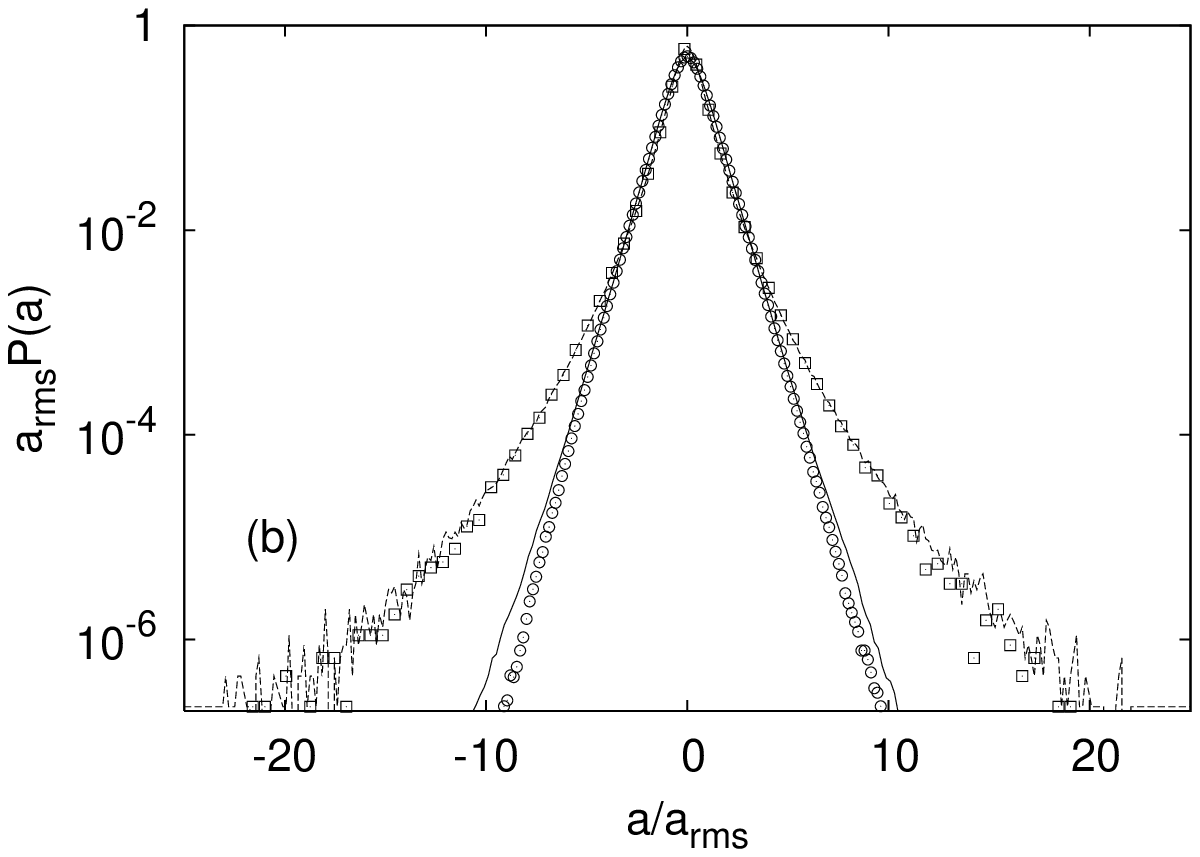}
\caption{(a) Acceleration pdf's for a subset of Stokes values
  ($St=0,0.16,0.37,0.58,1.01,2.03,3.31$ from top to bottom) at
  $R_{\lambda}=185$.  The inset displays the acceleration flatness,
  $\langle \bm a^4\rangle/\langle \bm a^2\rangle^2$, at increasing
  $R_\lambda$ from bottom to top.  (b) The two more external curves
  correspond to the acceleration pdf for $St=0.16$ ($\Box$) and the
  pdf of the fluid tracers acceleration measured on the same position
  of the inertial particles, ${D{\bm u}\over Dt}$ (solid line).  The
  two inner curves are the acceleration pdf at the highest Stokes,
  $St=3.31$, ($\circ$) and the pdf of the filtered fluid acceleration
  (solid line).  All curves are normalised to have unit variance.  }
\label{fig:2}
\end{figure}

Another interesting aspect shown in Fig.~\ref{fig:1}a is the residual
dependence of the normalised particle acceleration on Reynolds number.
For the case of fluid tracers it is known that intermittent
corrections to the dimensional estimate $a_{\rm rms} = a_0
(\epsilon^3/\nu)^{1/4}$ may explain the Reynolds dependence
(\cite{sawford,hill,biferale2}). Data suggest that the fluid
intermittency may be responsible of such deviations at $St>0$ as well.
This view is supported by the fact that the curves for the three
Reynolds numbers are almost parallel.

A two-parameters formula for the variance of the acceleration as a
function of Stokes number can be derived in the limit of vanishing
Stokes numbers as: $a_{\rm rms}^2(St) =a_{\rm rms}^2(0)+ C
\exp{[-(D/St)^{\delta}]}$ (\cite{falkprivate}). This expression
follows from the acceleration pdf of tracer particles under the
assumptions that (i) the main effect of inertia is to reduce the
particle concentration in regions where the accelerations is larger
than $\nu^{1/2}/{\tau_s}^{3/2}$; (ii) the pdf tail is well reproduced
by a stretched exponential shape with exponent $\beta = 2/3
\delta$. Although the formula fits well the data, the limitation of
our data-set to only a few points with $St \ll 1$ does not permit a
significant benchmark of the model.

In table \ref{table:2} we summarise the values that we have measured
for $\langle \bm a^2 \rangle $ and $\langle \bm a^4 \rangle$ as a
function of all Stokes and for all Reynolds numbers available.
Besides the effect of inertia on typical particle accelerations it is
also interesting to investigate the effects on the form of the
probability distribution function ${\bm a}(t)$. As shown in
Fig.~\ref{fig:2}a, the pdf's get less and less intermittent as $St$
increases. In the inset of the same figure we show the flatness,
$\langle \bm a^4 \rangle/ \langle \bm a^2 \rangle^2$, as a function of
$St$. The abrupt decreasing for $St>0$ is even more evident here
(notice that the $y$ scale is logarithmic).

 In the limits of small and large $St$ the qualitative trend of the
pdf's can be captured by the same arguments used for $a_{\rm rms}$.
In Fig.~\ref{fig:2}b we compare the pdf shape for the smallest Stokes
number with the one obtained by using the tracer acceleration measured
on the particle position, ${ D{\bm u}\over Dt}$.  As one can see the
two functions overlap perfectly, confirming that the only difference
between fluid particles and inertia particles for small Stokes is due
to preferential concentration. In the same figure we also compare for
the highest Stokes, $St=3.31$, the pdf of the particle acceleration
with the one obtained from the filtered fluid trajectories. Now the
agreement is less perfect but still fairly good, reassuring that this
limit can be captured starting from a low-pass filter of fluid tracer
velocities. It is worth mentioning that the pdf of tracer acceleration
measured on the particle position, ${ D{\bm u}\over Dt}$, approaches 
the unconditioned pdf as $St$ increases (not shown). This further
confirms that preferential concentration has a minor role on the
acceleration at these large Stokes values.

\section{Statistics of acceleration conditioned on the flow topology}
\label{sec:cond}
\begin{figure}
  \centerline{\includegraphics[draft=false,scale=0.54]{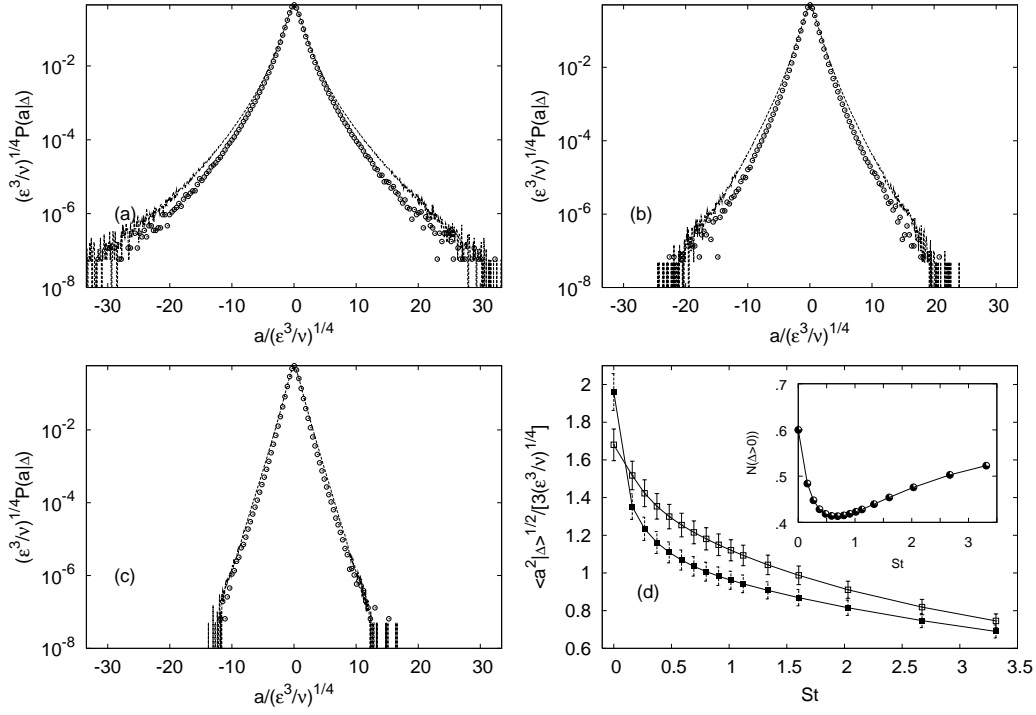}}
  \caption{Acceleration statistics conditioned on the sign of the
    discriminant $\Delta$ defined in (\protect\ref{eq:criteria}).  (a)
    Pdf of acceleration for $St=0.16$ conditioned on strain regions
    (solid line, $\Delta\leq 0$) and on rotating regions (points,
    $\Delta>0$) regions, respectively. (b) and (c) same as (a) for
    $St=0.48$ and $St=1.34$. (d) Normalised root mean square
    conditional acceleration on $\Delta\leq 0$ (empty boxes) and
    $\Delta>0$ (full boxes) regions as a function of $St$.  The inset
    displays the fraction of particles in the rotating regions
    $N(\Delta>0)\; (N(\Delta\leq 0)=1-N(\Delta>0))$ as a function of
    $St$. The conditional acceleration was computed on the data
    recorded at frequency $10 \tau_{\eta}$  (see table
    \protect\ref{table:1}). For $St=0$ the
    acceleration $\sqrt{\langle a^2 |\Delta\rangle/3}$ is estimated by
    using the pressure gradient $-\bm \nabla p$. }
\label{fig:3}
\end{figure}

We now focus on particle acceleration statistics conditioned on the
topological properties of carrier flow at particle positions.  In
particular, we look at the sign of the discriminant (see
e.g. \cite{chong} and \cite{Bec}):
\begin{equation}
\Delta=\left(\frac{{\rm det}[\hat{\sigma}]}{2} \right)^2-\left(
\frac{{\rm Tr}[\hat{\sigma}^2]}{6}\right)^3\,, \label{eq:criteria}
\end{equation}
being $\hat{\sigma}_{ij}=\partial_i u_j$ the strain matrix evaluated
at the particle position $\bm X$. Note that, in deriving
(\ref{eq:criteria}), we omitted the term proportional to  ${\rm
Tr}[\hat{\sigma}]$ because of incompressibility.  For $\Delta\leq 0$ the
strain matrix has $3$ real eigenvalues (strain dominated regions), for
$\Delta>0$ it has a real eigenvalue and $2$ complex conjugate ones
(rotational regions).  For a similar study, using a different 
 different characterisation of the flow
structures, see also  (\cite{SE91}). Note that in two-dimension the equivalent of $\Delta$ is
the well known Okubo-Weiss parameter that discerns elliptic from
hyperbolic regions of the flow.

In Figs.~\ref{fig:3}a,b,c we show the acceleration pdf, $P(a|\Delta)$,
conditioned on the sign of $\Delta$ at particle positions, for three
different characteristic Stokes numbers $St=0.16,0.48,1.34$. In
Fig~\ref{fig:3}d we show the root mean squared acceleration,
$\sqrt{\langle a^2 |\Delta\rangle/3}$, as a function of $St$.  A few
results are worth to be commented. The fraction of particles in the
two regions ($N(\Delta \gtrless 0)$) varies considerably as a function
of the Stokes number (see inset of Fig.~\ref{fig:3}d), with a
depletion of particles in the regions with some degree of rotation, which
however becomes less effective at large $St$.  This is similar to what
is observed in the inset of Fig.~\ref{fig:1}a, where the
non-homogeneous particle distribution is characterised in terms of the
correlation dimension (\cite{bccm05}). Further, despite the shape of
the pdf for a given Stokes number does not change much as a function
of the sign of $\Delta$, a noticeable change in the squared
acceleration is observed. As shown in Fig.~\ref{fig:3}d, the
acceleration is higher in the strain dominated regions than in the
 ones with some degree of rotation. We remark that the effect of inertia is
dramatic: for the smallest $St$ the conditional acceleration is larger when $\Delta <0$ while the opposite
behaviour is observed for tracer  ($St=0$).  This may be the
signature of the expulsion of particles out of intense vortex
filaments (which is more effective for $St\ll 1$) leading to an
undersampling of the acceleration in the regions dominated by
rotational motion.  The same difference is also measured for higher
moments of the conditioned acceleration (not shown).

These results point out that the strong correlation between flow
structure and particle preferential concentration is more effective at
low Stokes numbers. At larger $St$ the particle fraction $N(\Delta
\gtrless 0)$ approaches the tracer value (the response time is too
large to maitain the correlation between particle trajectories and the
local flow topology) and the depletion of acceleration should be
ascribed to the effect of filtering, as discussed in the previous
section ({\it cfr.} Fig.~\ref{fig:1}b and \ref{fig:2}b).
\section{Conclusions and perspectives}
A systematic study of the acceleration statistics of heavy particles
in turbulent flows, at changing both Stokes and Reynolds numbers has
been presented.  The main conclusions are (i) preferential
concentration plays an {\it almost singular} role at small
Stokes. Indeed, even a quite small inertia may suffice to expel
particles from those turbulent regions (vortex cores) where the most
intermittent and strong acceleration fluctuations would have been
experienced; (ii) for small Stokes, a good quantitative agreement
between the inertial particle acceleration and the conditioned fluid
tracer acceleration is obtained; (iii) at large Stokes, the main
effects is filtering of the velocity induced by the response Stokes
times. For $St >1 $, the statistical properties of fluid tracers
averaged over a time window of the order of $\tau_s$ are in a quite
good agreement with the inertial particle properties.

Some important questions remain open. 

It is not clear how to build up a phenomenological model that is able
to describe the inertial particles acceleration as a function of both
Stokes and Reynolds numbers.  For example, a naive generalisation of
the multifractal description, successfully used for fluid tracers
(\cite{biferale2}), may be insufficient. In fact, it is not
straightforward to include in such models the correlation between
preferential concentration and the local topological properties of the
carrier flow. Here such correlations have been studied in terms of the
real or complex nature of the eigenvalues of the strain matrix at
particle positions.  We found that, more effectively at small $St$
with respect to larger $St$ values, particles preferentially
concentrate in strain dominated regions and that this preferential
concentration has a clear role in determining the acceleration
fluctuations.  However, this information does not directly bring to a
model for the acceleration statistics.

The strong fluctuations of both Kolmogorov time and Kolmogorov
dissipative scale are certainly the most interesting aspects which
distinguish the statistics of heavy particles in turbulence from the
one measured in smooth flows.  It would be then important to study
also the statistical properties conditioned to the local Stokes number
(defined in terms of a ``local'' energy dissipation, see
e.g. \cite{Collins1}) .

Work in this direction will be reported elsewhere.

\begin{acknowledgments}
  We acknowledge useful discussions with G. Falkovich, E. Bodenschatz
  and Z. Warhaft. This work has been partially supported by the EU
  under the research training network HPRN-CT-2002-00300 ``Stirring
  and Mixing''. Numerical simulations have been performed thanks to
  the support of CINECA (Italy) and IDRIS (France) under the
  HPC-Europa project.  We thank also the ``Centro Ricerche e Studi
  Enrico Fermi'' and N.~Tantalo for support on the numerical
  computations.
\end{acknowledgments}


\begin{thebibliography}{} 
\bibitem[Balkovsky, Falkovich \& Fouxon 2001]{Falkovich-clustering}
  {\sc Balkovsky, E. Falkovich, G. \& Fouxon, A.} 2001 {Intermittent
    distribution of inertial particles in turbulent flows.} {\it
    Phys. Rev. Lett.}~{\bf 86}, 2790--2793.

\bibitem[Bec 2005]{Bec} {\sc Bec, J.} 2005 {Multifractal
concentrations of inertial particles in smooth random flows.}~{\it
J. Fluid Mech.}~{\bf 528}, 255-277.

\bibitem[Bec {\it et al.} 2005]{bccm05} {\sc Bec, J. Celani, A.
    Cencini, M. \& Musacchio, S.} 2005 {Clustering and collisions of
    heavy particles in random smooth flows.}~{\it Phys. Fluids}~{\bf
    17}, 073301.

\bibitem[Bec, Gawedzki \& Horvai 2004]{Bec-Gaw-Horvai} {\sc Bec,
    J. Gawedzki, K. \& Horvai, P.} 2004 {Multifractal clustering in
    compressible flows.}  {\it Phys. Rev. Lett.}~{\bf 92}, 224501.

\bibitem[Biferale {\it et al.} 2004]{biferale2} {\sc Biferale,
    L. Boffetta, G. Celani, A. Devenish, B.J. Lanotte, A. \& Toschi,
    F.} 2004 {Multifractal statistics of Lagrangian velocity and
    acceleration in turbulence.}  {\it Phys. Rev. Lett.}~{\bf 93},
  064502.

\bibitem[Biferale {\it et al.} 2005]{bife} {\sc Biferale, L. Boffetta,
    G.  Celani, A. Lanotte, A. \& Toschi, F.} 2005 {Particle trapping
    in three-dimensional fully developed turbulence.}  {\it
    Phys. Fluids.}~{\bf 17}, 021701.

\bibitem[Boffetta, De Lillo \& Gamba 2004]{Boffetta} {\sc Boffetta,
    G. De Lillo, F. \& Gamba, A.} 2004 {Large scale inhomogeneity of
    inertial particles in turbulent flows.}  {\it Phys. Fluids}~{\bf
    16}, L20--L24.

\bibitem[Boivin, Simonin \& Squires 1998]{Squires} {\sc Boivin,
    M. Simonin, O. \& Squires, KD.} 1998 {Direct numerical simulation of
    turbulence modulation by particles in isotropic turbulence.}  {\it
    J. Fluid Mech.}~{\bf 375}, 235--263.

\bibitem[Chen {\it et al.} 1993]{She} {\sc Chen, S.  Doolen, G.D.
    Kraichnan, R.H. \& She, Z.S.}  1993 {On statistical correlations
    between velocity increments and locally averaged dissipation in
    homogeneous turbulence.}~{\it Phys. Fluids} A~{\bf 5}, 458-463.

\bibitem[Chong, Perry \& Cantwell 1990] {chong} {\sc Chong,
M.S., Perry, A.E. \& Cantwell, B.J.} 1990 {A general classification
of three-dimensional flow field.} {\it Phys. Fluids A}~{\bf 2},
765--777.

\bibitem[Chun {\it et al.} 2005]{Collins2} {\sc Chun, J. Koch,
    D.L. Rani, S. Ahluwalia, A. \& Collins, L.R.} 2005  {Clustering of
    aerosol particles in isotropic turbulence.}  {\it J. Fluid
    Mech.}~{\bf 536}, 219--251.

\bibitem[Collins \& Keswani 2004]{Collins1} {\sc Collins, L.R. \& Keswani,
    A.} 2004 {Reynolds number scaling of particle clustering in
    turbulent aerosols.} {\it New J. Phys.}~{\bf 6}, 119.

\bibitem[Csanady 1980]{11} {\sc Csanady, G.} 1980 {\it Turbulent
    diffusion in the environment.} Geophysics and Astrophysics
  Monographs Vol. 3 D. Reidel Publishing Company.

\bibitem[Eaton \& Fessler 1994]{3} {\sc Eaton, J.K.\& Fessler, J.R.} 1994
  {Preferential concentrations of particles by turbulence.}  {\it
    Int. J.  Multiphase Flow}~{\bf 20}, 169--209.

\bibitem[Falkovich, Fouxon \& Stepanov 2002]{7} {\sc Falkovich, G.
    Fouxon, A. \& Stepanov, M.} 2002 {Acceleration of rain initiation by
    cloud turbulence.}  {\it Nature}~{\bf 419}, 151--154.

\bibitem[Falkovich \& Pumir 2004]{Falkovich-Pumir} {\sc Falkovich,
    G.\& Pumir, A.} 2004 {Intermittent distribution of heavy particles
    in a turbulent flow.}  {\it Phys.  Fluids}~{\bf 16}, L47--L51.

\bibitem[Falkovich 2005]{falkprivate} {\sc Falkovich, G.}  2005
  Private Communication.

\bibitem[Hill 2002]{hill} {\sc Hill, R.J.} 2002 {Scaling of
    acceleration in locally isotropic turbulence.}  {\it J. Fluid
    Mech.}~{\bf 452}, 361--370.
 
\bibitem[La Porta {\it et al.} 2001]{boden} {\sc La Porta, A. Voth,
    G.A. Crawford, A.M.  Alexander, J. \& Bodenschatz, E.} 2001 {Fluid
    particle accelerations in fully developed turbulence.} {\it
    Nature}~{\bf 409}, 1017--1019.

\bibitem[La Porta {\it et al.} 2002]{boden1} {\sc La Porta, A. Voth,
    G.A. Crawford, A.M.  Alexander, J. \& Bodenschatz, E.} 2002
  {Measurement of particle accelerations in fully developed
    turbulence.}  {\it J. Fluid Mech.}~{\bf 469}, 121--160.

\bibitem[Lewis \& Pedley 2000]{10} {\sc Lewis, D. \& Pedley, T.}  2000
  {Planktonic contact rates in homogeneous isotropic turbulence:
    Theoretical predictions and kinematic simulations.} {\it
    J. Theor. Biol.}~{\bf 205}, 377--408.

\bibitem[Maxey \& Riley 1983]{maxey} {\sc Maxey, M.R. \& Riley, J.}
  1983 {Equation of motion of a small rigid sphere in a nonuniform
    flow.}  {\it Phys. Fluids}~{\bf 26}, 883--889.

\bibitem[Mordant {\it et al} 2001]{Pinton} {\sc Mordant, N.  Metz, P.
    Michel, O. \& Pinton, J.-P.} 2001 {Measurement of Lagrangian
    velocity in fully developed turbulence.}  {\it
    Phys. Rev. Lett.}~{\bf 87}, 214501.

\bibitem[Pinsky \& Khain 1997]{5} {\sc Pinsky, M. \& Khain, A.} 1997
  {Turbulence effects on droplet growth and size distribution in
    clouds--a review.}  {\it J. Aerosol Sci.}~{\bf 28}, 1177--1214.
 
\bibitem[Post \& Abraham 2002]{4} {\sc Post, S. \& Abraham, J.}  2002
  {Modeling the outcome of drop-drop collisions in Diesel sprays.}
  {\it Int. J. of Multiphase Flow}~{\bf 28}, 997--1019.

\bibitem[Reade \& Collins 2000]{Collins} {\sc Reade, W.C. \& Collins,
    L.R.} 2000 {A numerical study of the particle size distribution of
    an aerosol undergoing turbulent coagulation.}  {\it J. Fluid
    Mech.}~{\bf 415}, 45--64.

\bibitem[Rothschild \& Osborn 1988]{9}{\sc Rothschild, B.J. \& Osborn,
    T.R.} 1988 { Small-scale turbulence and plankton contact rates.}
  {\it J. Plankton Res.}~{\bf 10}, 465--474.

\bibitem[Sawford \& Guest 1991]{pdfmod}{\sc Sawford, B.L. \& Guest,
    F.M.} 1991 {Lagrangian statical simulation of the turbulent motion
    of heavy particles.} {\it Boundary-Layer Meteorol.}~{\bf 54},
  147--166.

\bibitem[Sawford {\it et al.} 2003]{sawford} {\sc Sawford, B.L. Yeung,
    P.K. Borgas, M.S. Vedula, P.  La Porta, A. Crawford,
    A.M. Bodenschatz, E. } 2003 {Conditional and unconditional
    acceleration statistics in turbulence.}  {\it Phys. Fluids}~{\bf
    15}, 3478--3489.

\bibitem[Seinfeld 1986]{12} {\sc Seinfeld J.} 1986 {\it Atmospheric
    chemistry and physics of air pollution.} J. Wiley and Sons.

\bibitem[Shaw 2003]{8} {\sc Shaw, R.A.} 2003 {Particle-turbulence
    interactions in atmospheric clouds.}  {\it Ann. Rev. Fluid
    Mech.}~{\bf 35}, 183--227.


\bibitem[Squires \& Eaton 1991] {SE91} {\sc Squires K.D. \&
Eaton J.K.} 1991 {Preferential concentration of particles by
turbulence.} {\it Phys. Fluids A}~{\bf 3}, 1169--1178.

\bibitem[Villedieu \& Hylkema 2000]{prop} {\sc Villedieu P. \& Hylkema J.}
2000 {Mod\`eles num\'eriques lagrangiens pour la phase dispers\'ee
dans les propulseurs \`a poudre.} Rapport technique ONERA.

\bibitem[Warhaft, Gylfason, \& Ayyalasomayajula 2005]{wara} {\sc
    Warhaft, Z. Gylfason, A. \& Ayyalasomayajula, S.} 2005 private
  communication.

\bibitem[Zaichik, Simonin \& Alipchenkov 2003]{simo} {\sc Zaichik, L.I.
    Simonin, O. \& Alipchenkov V.M.} 2003 {Two statistical models for
    predicting collision rates of inertial particles in homogeneous
    isotropic turbulence.} {\it Phys. Fluids}~{\bf 15}, 2995--3005.

\bibitem[Zhou, Wexler \& Wang 2001]{Wang} {\sc Zhou, Y. Wexler, A. \&
    Wang, L.-P.}  {Modelling turbulent collision of bidisperse
    inertial particles.}  {\it J. Fluid Mech.}~{\bf 433}, 77--104.
\end{thebibliography}
\end{document}